\documentclass[useAMS,usenatbib]{mn2e}
\usepackage{epsfig}
\usepackage{amsmath}
\usepackage{rotating}
\usepackage{color}

\newcommand{\be}{\begin{equation}}
\newcommand{\beq}{\begin{equation}}
\newcommand{\ba}{\begin{eqnarray}}
\newcommand{\ee}{\end{equation}}
\newcommand{\eeq}{\end{equation}}
\newcommand{\ea}{\end{eqnarray}}

\newcommand{\hs}{\hspace{1mm}}

\newcommand{\apj}{ApJ}
\newcommand{\aap}{A\&A}
\newcommand{\apjl}{ApJL}
\newcommand{\mnras}{MNRAS}
\newcommand{\aj}{AJ}
\newcommand{\apjs}{ApJS}
\newcommand{\nat}{{\it Nature}}
\newcommand{\araa}{ARA\&A}
\newcommand{\pasj}{PASJ}

\def\lsim{~\rlap{$<$}{\lower 1.0ex\hbox{$\sim$}}}

\def\gsim{~\rlap{$>$}{\lower 1.0ex\hbox{$\sim$}}}

\title[Constraining the Ly$\alpha$ Escape Fraction]{Empirical Constraints on the Star Formation \& Redshift Dependence of the Ly$\alpha$ `Effective' Escape Fraction}

\author[Dijkstra \& Jeeson-Daniel]{Mark
Dijkstra$^{1}$\thanks{E-mail:dijkstra@mpa-garching.mpg.de} and Akila Jeeson-Daniel$^{1,2}$\\ $^{1}$Max Planck Institute for Astrophysics, Karl-Schwarzschild-Str. 1, 85741, Garching, Germany\\ $^{2}$School of Physics,
University of Melbourne, Parkville, Victoria, 3010, Australia}
\begin{document}

\date{\today}
\pagerange{\pageref{firstpage}--\pageref{lastpage}} \pubyear{2012}

\maketitle

\label{firstpage}
\begin{abstract}
We derive empirical constraints on the volume averaged `effective' escape fraction of Ly$\alpha$ photons from star forming galaxies as a function of redshift, by comparing star formation functions inferred directly from observations, to observed Ly$\alpha$ luminosity functions. Our analysis shows that the effective escape fraction increases from $f_{\rm esc}^{\rm eff} \sim 1-5\%$ at $z=0$, to $f_{\rm esc}^{\rm eff}\sim 10\%$ at $z=3-4$, and to $f_{\rm esc}^{\rm eff}=30-50 \%$ at $z=6$. Our constraint at $z=6$ lies above predictions by models that do not include winds, and therefore hints at the importance of winds in the Ly$\alpha$ transfer process (even) at this redshift.
We can reproduce Ly$\alpha$ luminosity functions with an $f_{\rm esc}^{\rm eff}$ that does not depend on the galaxies star formation rates ($\psi$) over up to $\sim 2$ orders of magnitude in Ly$\alpha$ luminosity. It is possible to reproduce the luminosity functions with an $f_{\rm esc}^{\rm eff}$ that decreases with $\psi$ - which appears favored by observations of drop-out galaxies - in models which include a large scatter ($\sigma \gsim 1.0$ dex) in $f_{\rm esc}^{\rm eff}$, and/or in which star forming galaxies only have a non-zero $f_{\rm esc}^{\rm eff}$ for a fraction of their life-time or a fraction of sightlines.
We provide a fitting formula that summarizes our findings. 
\end{abstract}

 
\section{Introduction}
\label{sec:intro}

The Ly$\alpha$ emission line is one of the most prominent features in the intrinsic spectrum of star forming galaxies \citep[e.g.][]{PP67,Schaerer03,J10}. The presence of a luminous, redshifted Ly$\alpha$ line has been used to spectroscopically confirm - and find -  galaxies out to $z\sim 7$ \citep[e.g.][]{Iye,Ota,Rhoads12}. 

Ly$\alpha$ emitting galaxies (LAEs hereafter)\footnote{In this letter, we use the term LAE to describe any Ly$\alpha$ emitting galaxy. It is also common in the literature to define LAEs as only those Ly$\alpha$ emitting galaxies that have been selected on the basis of their strong Ly$\alpha$ emission line.} are of interest for various reasons, including for example: ({\it i}) their continua are typically fainter than - and hence complement samples of- continuum selected (i.e. drop-out selected) galaxies; ({\it ii}) LAEs at $z>5$ provide an independent probe of the reionization epoch, as the Ly$\alpha$ line is affected by neutral intergalactic gas \citep[e.g.][]{HS99,MR04}; ({\it iii}) as Ly$\alpha$ photons likely scatter through the interstellar media (ISM) of galaxies, the total distance they travel through the ISM is enhanced compared to that of continuum photons. Ly$\alpha$ photons are therefore thought to provide a sensitive probe of the dust content (and gas kinematics) of the ISM; ({\it iv}) Ly$\alpha$ selected galaxies will be used to probe the equation of state of the dark energy at $z=1.9-3.5$ by the HETDEX\footnote{http://www.hetdex.org} experiment \citep{Hill08}.

The main uncertainty that affects interpretations of Ly$\alpha$ observations of LAEs relates to the complex radiative transfer of Ly$\alpha$ photons through both the ISM, the circum galactic medium (CGM), and intergalactic medium \citep[IGM, e.g.][]{Zheng10,DK12,Verhamme12,Laursen12,Cantalupo,Akila}. Moreover, these processes are not independent: radiative transfer at the ISM-level affects how the radiative transfer proceeds at the intergalactic level\footnote{The simplest way to illustrate this dependence is by considering that scattering of Ly$\alpha$ photons through outflows of HI gas (on - say - kpc scales) results in an overall redshift of the Ly$\alpha$ spectral line relative to other nebular lines \citep[e.g.][]{Zheng02,MD06,V08}. This overall redshift of the Ly$\alpha$ line reduces the probability that these photons subsequently scatter in the IGM \citep[][]{DW10}.}. In recent years Ly$\alpha$ RT has been modeled on all these scales, usually by combining simulated galaxies with Ly$\alpha$ radiative transfer calculations \citep[e.g][]{T06,Laursen07,Barnes,Verhamme12,Yajima}. These calculations are extremely difficult to carry out from first principles (see Dijkstra \& Kramer 2012), and ultimately must be constrained by observations.

The goal of this paper is to provide empirical (i.e. based purely on observations) constraints on the dependence of the {\it effective escape fraction}\footnote{The term `effective escape fraction' was coined previously by Nagamine et al. (2010), and is often simply referred to as `escape fraction'. In \S~\ref{sec:fobsvsfesc} we argue why we caution against universal usage of the term escape fraction, and why it helps to distinguish between an escape fraction and an effective escape fraction.} 
of Ly$\alpha$ photons, $f_{\rm esc}^{\rm eff} \equiv L_{\alpha}/L_{\alpha,\rm int}$, where $L_{\alpha}$ ($L_{\alpha,\rm em}$) denotes the observed (intrinsic) Ly$\alpha$ luminosity. Our goal is to constrain $f_{\rm esc}^{\rm eff}$ as a function of redshift \citep[as in][]{Hayes11,Blanc}. Furthermore, we investigate whether $f_{\rm esc}^{\rm eff}$ depends on the star formation rate of galaxies. Previous works by \citet{Hayes11} and \citet{Blanc} constrained the volume averaged effective escape fraction $f_{\rm esc}^{\rm eff}$ by comparing the star formation rate density, $\dot{\rho}_*$, inferred from the {\it observed} Ly$\alpha$ luminosity density, to $\dot{\rho}_*$ inferred from other observations. This method is highly non-trivial, because Ly$\alpha$ observations only detect galaxies for which $f_{\rm esc}^{\rm eff}$ exceeds some star formation rate-dependent value (at very low star formation rates, the Ly$\alpha$ flux falls below the detection threshold even when all Ly$\alpha$ photons made it to the observer), and one must attempt to account for this. For example, \citet{Hayes11} use UV-luminosity functions of drop-out galaxies to estimate the appropriate value for $\dot{\rho}_*$ at $z\gsim 2.5$, and a significant part of their analysis is devoted to choosing the proper lower integration limit when integrating over the UV-luminosity function. Our method uses star formation rate functions to estimate $f_{\rm esc}^{\rm eff}$. We show that this allows for more direct constraints which circumvent the difficulties associated with choosing such integration limits.

The outline of this paper is as follows: We describe in \S~\ref{sec:formalism} how we combine observations of Ly$\alpha$ luminosity functions of LAEs with observations of star formation functions, to put constraints on the effective escape fraction of Ly$\alpha$ photons, $f_{\rm esc}^{\rm eff}$. We present our main results in \S~\ref{sec:results} before presenting our conclusions in \S~\ref{sec:conc}. 

\section{Empirical Constraints on the Ly$\alpha$ Effective Escape Fraction}
\label{sec:formalism}

Ly$\alpha$ photons - just as H$\alpha$ photons - are emitted following recombination events in HII regions, and closely trace ongoing star formation. The H$\alpha$ luminosity of a galaxy is related to its star formation rate, denoted with $\psi$, as $L_{{\rm H}\alpha}=1.2\times 10^{41}\times(\psi/[{\rm M}_{\odot}\hs {\rm yr}^{-1}])$ erg s$^{-1}$ \citep[][this conversion assumes a Salpeter IMF in the mass range 0.1-100 $M_{\odot}$]{K98}. The intrinsic Ly$\alpha$ luminosity, denoted with $L_{\alpha,{\rm int}}$, is $\sim 8\times$ larger than the H$\alpha$ luminosity (for case-B recombination and $T=10^4$ K, e.g. Hayes et al. 2011), and we have 
\begin{equation}
L_{\alpha,{\rm int}}=k\times \Big{(} \frac{\psi}{M_{\odot}\hs{\rm yr}^{-1}}\Big{)},
\label{eq:lyaem}
\end{equation} where $k=10^{42}$ erg s$^{-1}$. The factor $k$ can be higher (or lower) by a factor of $\sim 2$ depending on the assumed IMF, and/or stellar metallicity. In the extreme case of a top-heavy IMF and zero-metallicity stars, the factor $k$ can be as high as $k\approx 10$ \citep[][]{Raiter}. Our constraints on $f_{\rm esc}^{\rm eff}$ scale with our assumed $k$ as $f_{\rm esc}^{\rm eff} \propto k^{-1}$. The `observed Ly$\alpha$ luminosity', defined as the observed flux multiplied by $4\pi d^2_{\rm L}(z)$ ($d_{\rm L}(z)$ denotes the luminosity distance out to redshift $z$), is
\begin{equation}
L_{\alpha}=f_{\rm esc}^{\rm eff}(\psi,z)\times L_{\alpha,{\rm int}},
\label{eq:fobs}
\end{equation} where  $f_{\rm esc}^{\rm eff}(\psi,z)$ denotes the effective escape fraction of Ly$\alpha$ photons.

We constrain the parameter $f_{\rm esc}^{\rm eff}(\psi,z)$ by comparing observed Ly$\alpha$ luminosity functions to observationally inferred star formation functions: the Ly$\alpha$ luminosity function, denoted by $\frac{dn}{d\log L_{\alpha}}d\log L_{\alpha}$, measures the comoving number density of galaxies with (the logarithm of their) Ly$\alpha$ luminosities in the range $\log L_{\alpha} \pm d\log L_{\alpha}/2$. The star formation function, denoted with $\frac{dn}{d\log \psi}d\log \psi$ measures the comoving number density of galaxies that are forming stars at rate (whose logarithm is) in the range $\log \psi \pm d\log \psi/2$. We describe the star formation functions used in our analysis, and how we convert these into Ly$\alpha$ luminosity functions in \S~\ref{sec:lyalf}. This conversion depends on $f_{\rm esc}^{\rm eff}(\psi,z)$, and we use observed Ly$\alpha$ luminosity functions to obtain constraints in \S~\ref{sec:results}.

\subsection{Star Formation Functions}
\label{sec:lyalf}

Star formation functions can be described by Schechter functions:

\begin{equation}
\frac{dn}{d\log \psi}={\rm ln}\hs 10 \times \psi \times \frac{\Phi_*}{\psi_*}\Big{(} \frac{\psi}{\psi_*}\Big{)}^{-\alpha}\exp(-\psi/\psi_*).
\end{equation} We adopt the redshift dependent Schechter function parameters from Smit et al. (2012, their Table~3)\footnote{For the data at $z=0.35$ we use the values from \citet{Bell07}. For the data at $z=0.35$ \citet{Bell07} combine UV and MIR luminosity functions to construct their star formation functions.  For the higher redshift star formation functions, \citet{Smit12} construct star formation functions from UV LFs, combined with constraints on the slope of the rest-frame continuum $\beta$.}. Our results are insensitive to this choice (see \S~\ref{sec:modelunc}).

In the absence of scatter, there is a one-to-one relation between $\psi$ and $L_{\alpha}$. The Ly$\alpha$ luminosity functions then relate to the star formation functions as
\begin{equation}
\frac{d n}{d \log L_{\alpha}} = \frac{dn }{d\log \psi}\frac{d\log \psi}{d \log L_{\alpha}}=\frac{dn }{d\log \psi}\Big{|}_{\psi=L_{\alpha}/(kf_{\rm esc}^{\rm eff})},
\label{eq:lfno}
\end{equation} where in the last equality we used Eq~\ref{eq:lyaem} and Eq~\ref{eq:fobs}.

In reality we do not expect each galaxy that forms stars at some rate $\psi$ to have exactly the same $f_{\rm esc}^{\rm eff}$. It is therefore reasonable to study models in which we assume that there is a dispersion (or scatter) in $f_{\rm esc}^{\rm eff}$ at a fixed $\psi$. In the presence of scatter, we generally have
\begin{eqnarray}
\frac{d n}{d \log L_{\alpha}} = \int_{-\infty}^{\infty}d \log \psi\hs \frac{dn }{d\log \psi}P(\log L_{\alpha} | \log \psi)
\label{eq:lfwi}
\end{eqnarray} where $P(\log L_{\alpha}|\log \psi)d \log L_{\alpha}$ denotes the probability that a galaxy that is forming stars at a rate $\psi$ has a Ly$\alpha$ luminosity in the range $\log L_{\alpha} \pm d \log L_{\alpha}/2$. We assume that the effective escape fraction $f_{\rm esc}^{\rm eff}$ has a scatter\footnote{We assume for simplicity that this scatter is independent of star formation rate $\psi$. \citet{Garel12} have recently presented a model in which the scatter in $f_{\rm esc}^{\rm eff}$ increases with $\psi$.} that is described by a (truncated) log-normal distribution. That is, 
\begin{eqnarray}
P(\log L_{\alpha}|\log \psi)\equiv \frac{dP}{d\log L_{\alpha}}= \frac{dP}{d \log f_{\rm esc}^{\rm eff}}\Big{|}_{f_{\rm esc}^{\rm eff}=L_{\alpha}/(k\psi)}= \nonumber \\
=\left\{\begin{array}{ll} \frac{\mathcal{N}}{\sqrt{2 \pi} \sigma}\exp\Big{(} \frac{-(\log f_{\rm esc}^{\rm eff}-\langle \log f_{\rm esc}^{\rm eff} \rangle)^2}{2\sigma^2}\Big{)} & f_{\rm esc}^{\rm eff} \leq 1 \\ \nonumber
0 & f_{\rm esc}^{\rm eff}>1
     \end{array}\right.,
     \label{eq:fobspdf}
\end{eqnarray} where $\mathcal{N}$ denotes a factor that ensures that the function $dP/d \log f_{\rm esc}^{\rm eff}$ is normalized.

\begin{figure*}
\vbox{\centerline{\epsfig{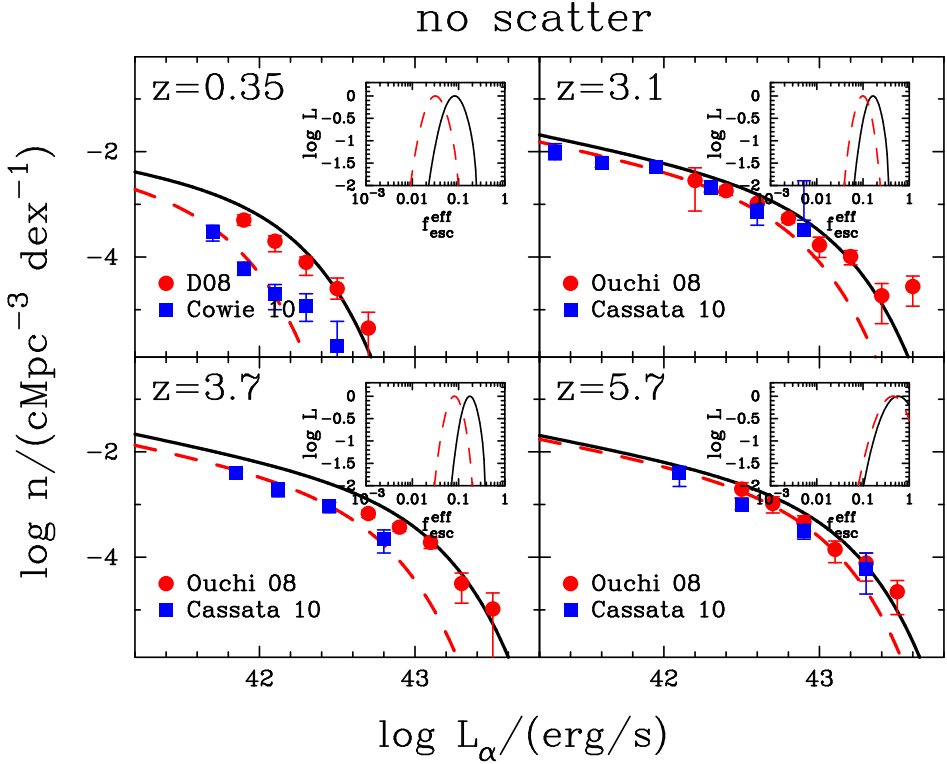}}}
\caption[]{The Figure compares observed Ly$\alpha$ luminosity functions (indicated by the data points) of LAEs at $z=0.35$ ({\it upper left}), $z=3.1$ ({\it upper right}), $z=3.7$ ({\it lower left}), and $z=5.7$ ({\it lower right}) with model predictions under the assumption that there is a one-to-one relation between star formation rate $\psi$ and observed Ly$\alpha$ luminosity $L_{\alpha}$ (see Eq~\ref{eq:lfno}, i.e. there is no scatter in $f_{\rm esc}^{\rm eff}$), for the best-fit observed Ly$\alpha$ fraction, $f_{\rm esc}^{\rm eff}$,  (as {\it black solid lines} and {\it red dashed lines}, see text). The {\it insets} show the likelihood $\mathcal{L}[f_{\rm esc}^{\rm eff}]$ as a function of $f_{\rm esc}^{\rm eff}$. This Figure illustrates that the models reproduce the Ly$\alpha$ luminosity functions well, except at the bright end (which may be contaminated by low luminosity AGN, see Ouchi et al. 2008). It is worth pointing out that $f_{\rm esc}^{\rm eff}$ is independent of $\psi$ in our models.}
\label{fig:fig1}
\end{figure*} 
\begin{figure*}
\vbox{\centerline{\epsfig{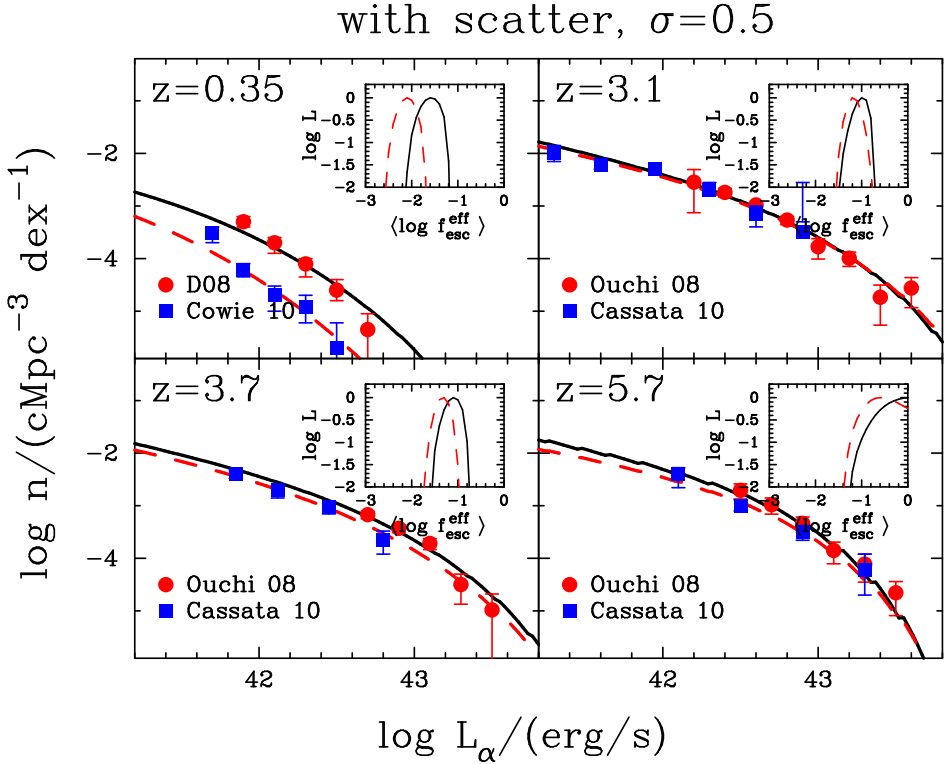}}}
\caption[]{Same as Figure~\ref{fig:fig1}, but for models in which we assume that there is a dispersion in $f_{\rm esc}^{\rm eff}$, described by a log-normal distribution with a standard deviation of $\sigma=0.5$, at a fixed $\psi$. This Figure shows that a dispersion in $f_{\rm esc}^{\rm eff}$ flattens the predicted Ly$\alpha$ luminosity functions, which improves the agreement with the data at $z=0.35$ and at high $L_{\alpha}$. These models obtain constraints from different datasets that agree better with each other.}
\label{fig:fig2}
\end{figure*} 

\subsection{Constraining the Effective Escape Fraction.}
\label{sec:fobs}
We first assume that $f_{\rm esc}^{\rm eff}(z,\psi)=f_{\rm esc}^{\rm eff}(z)$. That is, we first assume that $f_{\rm esc}^{\rm eff}$ is independent of the star formation rate $\psi$. We make this assumption because we will show later that the Ly$\alpha$ luminosity functions are - surprisingly - consistent with this assumption.

Our analysis focusses on the Ly$\alpha$ luminosity functions centered on $z=0.35$ from Deharveng et al. (2008, {\it red filled circles}) and Cowie et al. (2010, {\it blue filled squares}), $z=3.1$, $z=3.7$, and $z=5.7$ from Ouchi et al. (2008, {\it red filled circles}) and \citet[][{\it blue filled squares}]{C11}. At each redshift, we compute the posterior probability for a range of $f_{\rm esc}^{\rm eff}$ as $P(f_{\rm esc}^{\rm eff}) \propto \int d^3{\bf X}\hs \mathcal{L}[f_{\rm esc}^{\rm eff}]P(f_{\rm esc}^{\rm eff})P_{\rm s}({\bf X})$, where $\mathcal{L}[f_{\rm esc}^{\rm eff}]=\exp[-0.5\chi^2]$ denotes the likelihood, in which $\chi^2=\sum_i^{N_{\rm data}}({\rm model}_i-{\rm data}_i)^2/\sigma^2_i$. The function $P(f_{\rm esc}^{\rm eff})\equiv 1$ denotes the prior probability distribution for $f_{\rm esc}^{\rm eff}$: i.e. we assume no prior knowledge of $f_{\rm esc}^{\rm eff}$. We stress however that our results do not depend on our choice of prior.

Finally, the vector ${\bf X}$ contains the three Schechter function parameters ${\bf X}^T=$($\alpha,\psi_*,\Phi_*$). The function $P_{\rm s}({\bf X})$ describes the prior probability for having any combination of parameters:  we assumed that $P_{\rm s}({\bf X})$ is a multivariate Gaussian, i.e. $P_{\rm s}({\bf X})=\mathcal{N}\exp\Big{[} -\frac{1}{2}({\bf X}-{\bf \mu}_{X})^T{\bf C}^{-1}({\bf X}-{\bf \mu}_{X})\Big{]}$, where $\mathcal{N}$ denotes the normalization factor. The vector ${\bf \mu}_{X}$ contains the best fit values for each of the parameters. The covariance matrix ${\bf C}$ contains the measured uncertainties on the parameters\footnote{The covariance matrix in this case is a $3\times 3$ matrix whose entries are given by $C_{ij}=\sigma_i\sigma_j\rho_{ij}$. Here $\sigma_i$ denotes the uncertainty on parameter '$i$', and $\rho_{ij}$ denotes the correlation coefficient between parameter $i$ and $j$, and obey $\rho_{ij}=\rho_{ji}$.  These correlation coefficients are generally not given. Following Dijkstra \& Wyithe (2012) we assumed that $\rho_{\alpha,\psi_*}=\rho_{\psi_*,\Phi_*}=\rho_{\alpha,\Phi_*}=0.9$ at each redshift. By definition $\rho_{ii}=1$ for all $'i'$. } (the most likely values and their uncertainties were taken from Smit et al. 2012).

\section{Results}
\label{sec:results}

\subsection{No Scatter}
\label{sec:noscat}
Figure~\ref{fig:fig1} shows four panels, each of which corresponds to one redshift bin. The observed Ly$\alpha$ luminosity functions that we used in our analysis are shown as the datapoints. The {\it inset} in each panel shows $\mathcal{L}[f_{\rm esc}^{\rm eff}]$ as a function of $f_{\rm esc}^{\rm eff}$. These panels contain two lines, both of which were obtained by fitting to a single data set. For example, we obtained the {\it black solid line} ({\it red dashed line}) in the $z=0.35$ panel by fitting to the data from \citet{D08} (Cowie et al. 2010)\footnote{The origin of the difference between the luminosity functions derived by \citet{D08} and \citet{Cowie10} appears to be in the incompleteness correction, which is large in \citet{D08}, but not in \citet{Cowie10}. \citet{Cowie10} note that this difference may be caused by a missing color correction in the Deharveng analysis (and quote private communication).}. At each redshift, we show the model luminosity functions for which both $\mathcal{L}[f_{\rm esc}^{\rm eff}]$-curves are maximized, using the same line color and style.

The {\it upper left panel} shows that the data from \citet{D08} translates to $f_{\rm esc}^{\rm eff}=8.5 \pm 3 \%$, while the data from Cowie et al. (2010) implies $f_{\rm esc}^{\rm eff}=3^{+2}_{-1} \%$ at $z=0.35$. Here the errorbars denote 68\% confidence levels, where we use the so-called `shortest interval' method \citep[see][and references therein]{Andrae} to determine the confidence intervals. The {\it upper right panel} shows that the effective escape fraction increases to $f_{\rm esc}^{\rm eff}=17 \pm 5\%$ [$f_{\rm esc}^{\rm eff}=10 \pm 3 \%$] for the Ouchi et al. (2008) [Cassata et al. (2010)] data at $z=3.1$, and to $f_{\rm esc}^{\rm eff}=17 ^{+6}_{-4}\%$ [$f_{\rm esc}^{\rm eff}=8 \pm 3\%$] for the Ouchi et al. (2008) [Cassata et al. 2010] data at $z=3.7$. Finally, we find that the effective escape fraction increases to $f_{\rm esc}^{\rm eff}=57 ^{+34}_{-21}\%$ [$f_{\rm esc}^{\rm eff}=44 ^{+28}_{-22}\%$] for the Ouchi et al. (2008) data [Cassata et al. (2010)] at $z=5.7$.

Our quoted uncertainties are statistical only, and do not take into account systematic uncertainties associated with the determination of observed Ly$\alpha$ luminosity functions. The different constraints we obtain on $f_{\rm esc}^{\rm eff}$ from different data-sets may reflect these systematic uncertainties: in particular, at $z\geq 3.1$ the data from Ouchi et al. (2008) derive from a narrow-band survey for LAEs, while the data from Cassata et al. (2010) derive from a deep spectroscopic survey (see \S~\ref{sec:modelunc} for a more detailed discussion of systematic uncertainties).

Figure~\ref{fig:fig1} shows that our models reproduce the individual datasets of observed Ly$\alpha$ luminosity functions well. Different datasets can result in slightly different constraints on $f_{\rm esc}^{\rm eff}$. It is striking that at $z \geq 3.1$ (especially $z=3.1$ and $z=5.7$), a $\psi$-independent $f_{\rm esc}^{\rm eff}$ reproduces the observations well over up to two orders Ly$\alpha$ luminosity, and therefore $\psi$. If anything, our models do not produce enough bright LAEs, which could be solved by having $f_{\rm esc}^{\rm eff}$ {\it increase} with $\psi$. Note however, that \citet{Ouchi08} point out that the bright end (i.e. at $\log L_{\alpha} \gsim 43.4$) may be contaminated by low luminosity AGN. The only data-set that we cannot reproduce well is that of \citet{Cowie10}: our model predicts significantly fewer LAEs than their two data-points at $\log L_{\alpha}\gsim 42.3$. This discrepancy could again be (partially) resolved by having $f_{\rm esc}^{\rm eff}$ increase with $\psi$. As we show below (in \S~\ref{sec:withscat}), we also significantly improve the agreement with the data when we introduce a scatter in $f_{\rm esc}^{\rm eff}$. 

\subsection{With Scatter}
\label{sec:withscat}

Figure~\ref{fig:fig2} shows the same as Figure~\ref{fig:fig1}, but here the models include a dispersion in $f_{\rm esc}^{\rm eff}$ (Eq~\ref{eq:lfwi}), which is described by a (truncated) log-normal distribution with a standard deviation of $\sigma=0.5$. This choice for $\sigma$ is a bit arbitrary, but can be justified by the work of \citet{DWes}, who found that the ratio of the Ly$\alpha$ to the UV-derived star formation rate can be described by a log-normal distribution with $\sigma=0.4$. We stress that changes to our main results are insignificant, even if we adopted $\sigma=1.0$.

Figure~\ref{fig:fig2} shows that a dispersion in $f_{\rm esc}^{\rm eff}$ flattens the predicted luminosity functions, and smoothens out the sharp-turnover in the predicted luminosity function. Both these changes help to improve the fit to the data at $z=0.35$ (and also at the highest $L_{\alpha}$ data point at $z=3.1$). Importantly, these models obtain constraints from different datasets that agree better with each other: for example, the best-fit models to the data from Ouchi et al. (2008) also provide decent fits to the data from Cassata et al. (2010).

In spite of the flattening of the predicted luminosity functions, these models still reproduce the data with an $\langle \log f_{\rm esc}^{\rm eff} \rangle$ that is independent of $\psi$. It is only when we adopt $\sigma\gsim 1.0$, that the predicted luminosity functions become flatter than the observations. Garel et al. (2012) have recently predicted that the scatter in $f_{\rm esc}^{\rm eff}$ increases with $\psi$, and that may be even larger than this at $\psi \gsim 20\hs M_{\odot}$ yr$^{-1}$. For models that include this large scatter, the data would require $\langle \log f_{\rm esc}^{\rm eff} \rangle$ to decrease with $\psi$. Such a requirement would be expected given observations of drop-out galaxies, which show evidence that the fraction of continuum selected galaxies that have `strong' Ly$\alpha$ emission lines increases towards lower UV-luminosities \citep[e.g.][]{Stark10}. This suggests that Ly$\alpha$ photons have an easier time escape from galaxies with lower UV luminosities, and therefore likely from galaxies with lower star formation rates $\psi$. 

The {\it insets} of Figure~\ref{fig:fig2} show $\mathcal{L}(\langle \log f_{\rm esc}^{\rm eff}\rangle)$. For example, the {\it inset} in the {\it upper right panel} shows that the best-fit $\langle \log f_{\rm esc}^{\rm eff} \rangle$ at $z=3.1$ is $10^{\langle \log f_{\rm esc}^{\rm eff} \rangle} \sim 0.06-0.1$, which lie a factor of $\sim 1.7$ below\footnote{At $z=5.7$ the best-fit $\langle \log f_{\rm esc}^{\rm eff} \rangle$ lies above the best-fit $f_{\rm esc}^{\rm eff}$ that we inferred in the absence of scatter.  This is because the expectation value, $E(f_{\rm esc}^{\rm eff})$, becomes {\it less} than $10^{\langle \log f_{\rm esc}^{\rm eff} \rangle}$ for $10^{\langle \log f_{\rm esc}^{\rm eff} \rangle} \gsim 0.3$ when $\sigma=0.5$ for the truncated PDFs that we assign to $f_{\rm esc}^{\rm eff}$.} the best-fit $f_{\rm esc}^{\rm eff}$ we derived for the model with no scatter. The best-fit values of $\langle \log f_{\rm esc}^{\rm eff}\rangle$ depend on the choice of $\sigma$: the larger $\sigma$, the smaller $\langle f_{\rm esc}^{\rm eff}\rangle$. The reason for this reduction is that  at fixed $\langle\log  f_{\rm esc}^{\rm eff}\rangle$, increasing $\sigma$ increases the {\it expectation value} $E(f_{\rm esc}^{\rm eff})$, which is given by $E(f_{\rm esc}^{\rm eff}) =\int_0^1 df_{\rm esc}^{\rm eff}\hs f_{\rm esc}^{\rm eff}\frac{dP}{df_{\rm esc}^{\rm eff}}$. We have verified that this best-fit expectation value $E(f_{\rm esc}^{\rm eff})$ barely depends on our choice of $\sigma$.  

We now practically have three `measures' of $f_{\rm esc}^{\rm eff}$ (namely $f_{\rm esc}^{\rm eff}$, $\langle \log f_{\rm esc}^{\rm eff} \rangle$, and $E(f_{\rm esc}^{\rm eff})$), which may be a bit confusing. We have therefore briefly summarized the meaning of these symbols in Table~\ref{table:symbols}.
\begin{table}
\caption{Summary of `Different' Measures of $f_{\rm esc}^{\rm eff}$.}  \centering.
\begin{tabular}{l l}
\hline\hline
symbol & meaning \\
\hline\\
  $f_{\rm esc}^{\rm eff}$ & effective escape fraction of Ly$\alpha$ of a galaxy\\
  $\langle \log f_{\rm esc}^{\rm eff} \rangle$ & mean of log $f_{\rm esc}^{\rm eff}$ in a  lognormal PDF (Eq~\ref{eq:fobspdf})\\
  $E(f_{\rm esc}^{\rm eff})$& expectation value of $f_{\rm esc}^{\rm eff}$ for models with\\
  & a lognormal PDF\\
 \hline\hline
\end{tabular}
\label{table:symbols}
\end{table}

\subsection{Comparing our $f_{\rm esc}^{\rm eff}(z)$ with Previous Works}
\begin{figure}
\vbox{\centerline{\epsfig{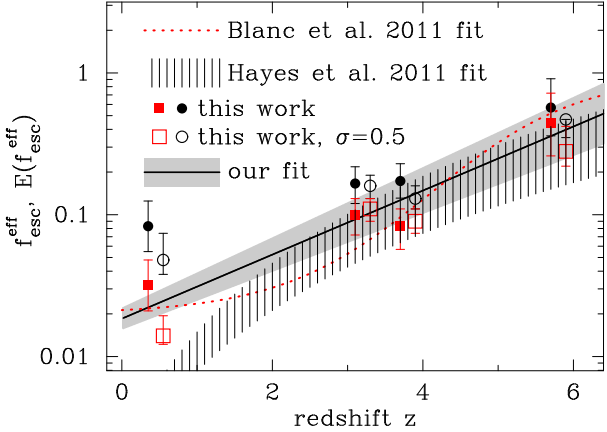}}}
\caption[]{This Figure compares our constraints on $f_{\rm esc}^{\rm eff}$ with the analytic fitting formula provided by Hayes et al. (2011, indicated by the {\it black shaded region}). The {\it filled symbols} (off-set by $\Delta z =0.2$ for clarity) represent the constraints on $f_{\rm esc}^{\rm eff}$ for the `no-scatter' models. The {\it open symbols} represent our constraints on the expectation values of $E(f_{\rm esc}^{\rm eff})$ for our models that include scatter. At each redshift, we have two data points which correspond to different data sets. The {\it grey region} represents our fitting formula (Eq~\ref{eq:fit}) and its uncertainties. Our work is consistent with Hayes et al. (2011), except at $z=0.35$ where our inferred $f_{\rm esc}^{\rm eff}$ is  higher, which is likely related to systematic uncertainties (see text). Our constraint at this redshift is in better agreement with that given by the `transition model' fit of Blanc et al. (2011, indicated by the {\it red dotted line}).}
\label{fig:fig3}
\end{figure}

We compare our inferred redshift evolution of $f_{\rm esc}^{\rm eff}(z)$ to the power-law fitting function from \citet{Hayes11} in Figure~\ref{fig:fig3}. 
The {\it filled symbols} represent the constraints on $f_{\rm esc}^{\rm eff}$ that we obtained for the `no-scatter' models in \S~\ref{sec:noscat}. The {\it open symbols} represent our constraints on the best-fit expectation value $E(f_{\rm esc}^{\rm eff})$ (these constraints do not depend on our adopted $\sigma$). We have off-set these data points by $\Delta z =0.2$ for clarity. At each redshift, we have two data points which correspond to different data sets. Including scatter reduces the expectation value of $f_{\rm esc}^{\rm eff}$ compared to models that have no scatter, typically by about $\sim 1\sigma$. The reduction is a bit larger at $z=0.35$. However, here we point out that the models that do not include scatter had difficulties fitting the data to begin with, and the constraints that we inferred from these models were likely less reliable.

Figure~\ref{fig:fig3} shows that our best-fit values are consistent with \citet{Hayes11} at $z\geq 3.1$, albeit on the high end of their quoted range. At $z=0.35$, our constraints on $f_{\rm esc}^{\rm eff}$ lie significantly higher than those of Hayes et al. (2011), who found $f_{\rm esc}^{\rm eff} =1.3 \pm 0.9 \%$ using the data from \citet{D08}, and $f_{\rm esc}^{\rm eff}=0.3 \pm 0.2 \%$ using the data from Cowie et al. (2010). These values would clearly not allow us to reconstruct the observed Ly$\alpha$ luminosity functions. Our constraint at $z=0.35$ is in better agreement with the `transition model' fit by Blanc et al. (2011). Their fit is represented by the {\it red dotted line} (which shows the `no LF integration limit' fit), which predicts $f_{\rm esc}^{\rm eff}\sim 2\%$ at $z=0.35$. This same fit gives slightly lower values for $f_{\rm esc}^{\rm eff}$ at $1\lsim  z  \lsim 3.5$ than ours.

A possible explanation for the lower preferred values for $f_{\rm esc}^{\rm eff}$ at $z=0.35$ by \citet{Hayes11} is that they compare the observed Ly$\alpha$ luminosity density to a total star formation rate density of $\dot{\rho}_*\sim 30\times 10^{-3}$ M$_{\odot}$ yr$^{-1}$ cMpc$^{-3}$. This value corresponds to the total integrated star formation rate density \citep[see Table~1 of][]{Bothwell11}. \citet{Bothwell11} show that galaxies with $\psi \gsim 10$ M$_{\odot}$ yr$^{-1}$ account only for $\sim 20\%$ of $\dot{\rho}_*$. If we consider an extreme example in which all galaxies with $\psi < 10$ M$_{\odot}$ yr$^{-1}$ have $f_{\rm esc}^{\rm eff}=3\%$, then their observed Ly$\alpha$ luminosity would be $L_{\alpha}\sim 3\times 10^{41}$ erg s$^{-1}$, which lies below the minimum detectable Ly$\alpha$ luminosity. The luminosity function presented by \citet{Cowie10} is therefore consistent with all galaxies $\psi < 10$ M$_{\odot}$ yr$^{-1}$ having $f_{\rm esc}^{\rm eff}=3\%$. Even if all galaxies with $\psi > 10$ M$_{\odot}$ yr$^{-1}$ would have $f_{\rm esc}^{\rm eff}=0$, then $f_{\rm esc}^{\rm eff}$ averaged over the population as a whole would be $\sim 2.4\%$, which is almost an order of magnitude higher than the value reported by \citet{Hayes11}. While the example we discussed here is clearly not realistic, it nevertheless shows that for very small $f_{\rm esc}^{\rm eff}$, large systematic uncertainties may be associated with estimating $f_{\rm esc}^{\rm eff}$ by comparing an observed Ly$\alpha$ luminosity density to a star formation rate density. Another way to phrase this is that for very small $f_{\rm esc}^{\rm eff}$, existing observations probe luminosities that are likely close to (or even larger than) $L_*$ in the Ly$\alpha$ luminosity function. In these cases, it is difficult and uncertain to estimate the Ly$\alpha$ luminosity density in faint, undetected sources.

We have also indicated a (ad-hoc) fitting formula that we found to capture the redshift evolution of our inferred $f_{\rm esc}^{\rm eff}$ reasonably well:
\begin{equation}
f_{\rm esc}^{\rm eff}(z,\psi)=\exp(-\tau_{\rm eff}),\hs\hs \tau_{\rm eff}=a_1+a_2z
\label{eq:fit}
\end{equation}, where $a_1=4.0\pm 0.16$ and $a_2=-0.52\pm0.05$. We obtain best fit values for $a_1$ and $a_2$ by minimizing $\chi^2$, which we compute using all 16 data points shown in Figure~\ref{fig:fig3}. The redshift evolution of $f_{\rm esc}^{\rm eff}$ for the best fit combination of $a_1$ and $a_2$ is represented by the {\it black solid line} in Figure~\ref{fig:fig2}. The upper/lower boundary of the {\it grey region} represents our fitting formula when we simultaneously subtract/add $\sigma/\sqrt{2}$ to both $a_1$ and $a_2$.  Here, uncertainties on the parameters $a_1$ and $a_2$ represent marginalized $1\sigma$ uncertainties. The fitting formula Eq.~\ref{eq:fit} captures our main results well, and further `predicts' that $f_{\rm esc}^{\rm eff}\sim 5^{+2}_{-1} \%$ at $z=2$, which is consistent with Hayes et al. (2010) who found $f_{\rm esc}^{\rm eff}=5.3 \pm 3.8 \%$ by comparing Ly$\alpha$ to H$\alpha$ luminosity functions. We have also applied our analysis to the more recent $z\sim 1$ data of Barger et al. (2012, not shown here), and found $f_{\rm esc}^{\rm eff}=5\pm 1 \%$ for the `no scatter case', which is also captured reasonably well by our fitting formula.

\section{Discussion}

\subsection{`Effective Escape' Fraction vs. `Escape' Fraction}
\label{sec:fobsvsfesc}

We explicitly differentiate between the term `escape' fraction and `effective escape' fraction, because these two quantities can take on very different values. In theoretical calculations that follow the transport of Ly$\alpha$ photons through a dusty medium, it is straightforward to compute the fraction of photons that are not absorbed by dust, and hence `escape' \citep[e.g.][]{Neufeld90,Hansen06,Laursen07,Yajima,Laursen12,Yajima13}. However, a large fraction of Ly$\alpha$ photons that escape from this medium can scatter in the surrounding circum-galactic and/or intergalactic medium and give rise to a low surface brightness Ly$\alpha$ glow around galaxies \citep[e.g.][]{IGM,Zheng10,Laursen11,Zheng11,Steidel11,Akila}. The surface brightness of this scattered radiation is typically much fainter than can be observed\footnote{For example, the surface brightness threshold for the $z=5.7$ LAE survey by Ouchi et al. (2008) is $\sim 10^{-18}$ erg s$^{-1}$ cm$^{-2}$ arcsec$^{-2}$. \citet{Rauch08} managed to go a factor of $\sim 10$ deeper in a $\sim 100$ hr exposure on the VLT.}, and this Ly$\alpha$ radiation would effectively be lost in observations. For example, \citet{Zheng10} find that Ly$\alpha$ scattering in the ionized IGM at $z=5.7$ rendered $80-95\%$ of all emitted Ly$\alpha$ radiation undetectable (consistent with the other studies, see \S~\ref{sec:modelcon}). There is no dust in their simulations, and the escape fraction of Ly$\alpha$ photons is $100 \%$. In contrast, the effective escape fraction would only be $f_{\rm esc}^{\rm eff} \sim 5-20\%$.

There is also observational evidence for the existence of spatially extended low surface brightness Ly$\alpha$ emission around galaxies  \citep[e.g.][]{Fynbo,Ostlin,Rauch08,Steidel11,Matsuda12,Hayes13}. \citet{Steidel11} detected spatially extended Ly$\alpha$ emission after stacking Ly$\alpha$ observations on 92 $z\sim 2.6$ LBGs, which allowed them to probe Ly$\alpha$ emission down to $\sim 10$ times fainter surface brightness levels. The total flux in their spatially extended halos significantly exceeded the total Ly$\alpha$ flux coming directly from their galaxies. The observations by \citet{Steidel11} imply that the escape fraction of Ly$\alpha$ photons can exceed the effective escape fraction significantly for surface brightness thresholds that are typical for current observations. Similarly, \citet{Matsuda12} detected Ly$\alpha$ halos around $z=3.1$ LAEs and found that the size of the halos (at fixed UV-luminosity of the LAEs) increases with local density (measured by the number density of LAEs). This dependence may help explain why other groups have not detected\footnote{Recently, \citet{Jiang} did not detect spatially extended Ly$\alpha$ emission around stacks of $43$ $z=5.7$ LAEs, and $40$ $z=6.5$ LAEs. At these redshifts there is room to hide a significant Ly$\alpha$ flux in the halo, even for the surface brightness threshold of $\sim 10^{-19}$ erg s$^{-1}$ cm$^{-2}$ arcsec$^{-2}$ that is reached in the stacking analysis. \citet{Jiang} comment that these observations indeed still appear broadly consistent with the predictions by \citet{Zheng11}.}
 spatially extended Ly$\alpha$ halos around LAEs (e.g. Feldmeier et al. 2013). In any case, the {\it possibility} that there is more Ly$\alpha$ flux in diffuse Ly$\alpha$ halos than in a compact source illustrates that the effective escape fraction - and previous determinations of this quantity - depend on the surface brightness threshold of the survey of interest (or the size of the photometric aperture in fixed aperture photometry), while the escape fraction does not \citep[also see][]{Yajima}.

The universal usage of the term escape fraction complicates comparisons between different studies: for example, \citet{Yajima13} compute true Ly$\alpha$ escape fractions in simulated galaxies as a function of redshift. Similarly, semi-analytic studies that model LAEs at $z=3-6$ \citep[e.g.][]{Kobayashi,Dayal11,Shimizu11,Jaime} introduce an escape fraction, which corresponds to a true escape fraction. Caution must be exercised when comparing these escape fractions to the observationally inferred effective escape fractions (as in Hayes et al. 2011, Blanc et al. 2011, and in this paper). Moreover, in some (but not all) studies the constraints on $f_{\rm esc}$ (and/or $f_{\rm esc}^{\rm eff}$) involve a `correction' for scattering in the IGM.  We stress that this correction is highly uncertain, as it depends on the radiative transfer at the interstellar and circum-galactic level (see \S~1).

\subsection{Comparison to Previous Works}
\label{sec:comp2}

We already compared our results to those obtained by Hayes et al. (2011, and also Blanc et al. 2010). Our approach, in which we use star formation functions and Ly$\alpha$ luminosity functions to constrain $f_{\rm esc}^{\rm eff}$, is similar to that adopted in theoretical studies. For example, \citet{LD06} use semi-analytic models - while e.g. \citet{Nagamine10} use hydrodynamical simulations - to generate star formation functions\footnote{To be precise, these models generate intrinsic Ly$\alpha$ luminosity functions, which give the number density of galaxies as a function of Ly$\alpha$ luminosity. This intrinsic Ly$\alpha$ luminosity function is practically the same as a star formation function.}, and then use Ly$\alpha$ luminosity functions to constrain $f_{\rm esc}^{\rm eff}$ at $z=3-6$. Importantly, the models that are used to generate the theoretical star formation functions are typically constrained by observations.  However, these (almost the same) observations can be converted directly into star formation functions, i.e. without generating the intermediate theoretical model. Indeed, our method completely circumvents this intermediate step. The fact that we can side-step this (substantial) part of the calculations allow us to more efficiently explore a larger suite of models for $f_{\rm esc}^{\rm eff}$, and to explore the impact of uncertainties with the observationally inferred star formation functions on our results.

Our results are broadly consistent with these previous theoretical studies: \citet{Nagamine10} find that $f_{\rm esc}^{\rm eff}=0.1$ at $z=3.1$, which is in excellent agreement with our results. \citet{Nagamine10} find $f_{\rm esc}^{\rm eff}=0.15$ at $z=6$, which is a factor of $\sim 2-3$ lower than what we find. The origin of this difference is unclear, but the {\it lower right panel} of Figure~\ref{fig:fig2} shows that the value preferred by \citet{Nagamine10} ($10^{\langle \log f_{\rm esc}^{\rm eff} \rangle} =0.15$) is not ruled out at great significance. \citet{LD06} find $f^{\rm eff}_{\rm esc} \sim 0.02$ at $z=3-6$. However, a redshift-dependent fraction of stars form in bursts with a top-heavy IMF for which $k\sim 10$ (see \S~\ref{sec:formalism}) in their models. Hayes et al. (2011) show that if this top-heavy IMF is replaced with a standard Salpeter IMF, that then the constraints obtained by \citet{LD06} agree well with \citet{Nagamine10} at $z=3-6$.

Finally, \citet{Nagamine10} showed that while their models with a constant $f_{\rm esc}^{\rm eff}$ fit the data well (in good agreement with our work), they obtain better fits using so-called `duty cycle' models, in which $\frac{dn}{d \log L_{\alpha}}=\epsilon_{\rm DC}\frac{dn}{d\log \psi}\big{|}_{\psi=L_{\alpha}/kf_{\rm esc}^{\rm eff}}$. These models represent a scenario in which star forming galaxies only have non-zero $f_{\rm esc}^{\rm eff}$ for a fraction $\epsilon_{\rm DC}$ of their lifetimes. We note that this may also represent a scenario in which Ly$\alpha$ escapes anisotropically from galaxies, and in which $f_{\rm esc}^{\rm eff}>0$ only along a fraction $\epsilon_{\rm DC}$ of the sightlines from them. The duty cycle parameter $\epsilon_{\rm DC}$ can also be incorporated in the $f_{\rm esc}^{\rm eff}$-PDF, simply by adding a Dirac-delta function at $f_{\rm esc}^{\rm eff}=0$ (after which we must renormalize the full-PDF). We have repeated our analysis including a duty cycle of $\epsilon_{\rm DC}=0.25$ into our $f_{\rm esc}^{\rm eff}$-PDF, and found that these models flatten the predicted luminosity functions, similarly to models with a non-zero scatter in $f_{\rm esc}^{\rm eff}$. These `duty-cycle model' therefore provide somewhat better fits to the luminosity functions (for models with $\sigma=0$), mostly because they improve the fits at the bright ends (just as our models with $\sigma=0.5$), in agreement with \citet{Nagamine10}. Moreover, the best-fit {\it expectation values} of $f_{\rm esc}^{\rm eff}$ in these duty cycle models\footnote{If we denote the expectation value of  $f_{\rm esc}^{\rm eff}$ along sightlines (or during time-intervals) where $f_{\rm esc}^{\rm eff}>0$ with $E(f_{\rm obs,DC})$. The overall expectation value is then given by $E(f_{\rm obs,DC})=\epsilon_{\rm DC}E(f_{\rm esc}^{\rm eff})$.} are consistent with our those obtained previously.

\subsection{Model Uncertainties}
\label{sec:modelunc}

A potential caveat is that (some of) our adopted Ly$\alpha$ luminosity functions were constructed from narrow-band surveys. Such surveys do not only impose a Ly$\alpha$ flux cut, but in practice also a cut in Ly$\alpha$ equivalent width (EW). For example, \citet{Ouchi08} adopt color-color criteria to select LAEs at $z=3.1$ that translate (roughly) to EW$\gsim 64$ \AA. We may worry that this data-set therefore misses a significant fraction of Ly$\alpha$ emitting galaxies. In practise however, the equivalent width cut does not appear to affect determinations of the Ly$\alpha$ luminosity functions: \citet{Gronwall07} present a luminosity function at $z=3.1$ that agrees well with \citet{Ouchi08}, even though they effectively apply a different EW-cut of EW$\gsim 20$ \AA. Moreover, Cassata et al. (2010) derived their luminosity functions from a spectroscopic survey, which does not employ any EW cut. Our inferred value for $f_{\rm esc}^{\rm eff}$ from the Cassata et al. (2010) data was in fact lower than that for the Ouchi et al. (2008) data, which suggests that uncertainties associated with how different LAE samples are constructed are subdominant to other systematic uncertainties.

Our analysis uses Schechter functions to describe the star formation rate functions. Recenty, \citet{Salim12} have demonstrated that a superior fit to star formation functions can be obtained from `Saunders' functions \citep[introduced by][]{Saunders}, given by
\begin{equation}
\frac{dn}{d\psi}=\frac{\Phi_*}{\psi_*}\Big{(} \frac{\psi}{\psi_*}\Big{)}^{-\alpha}\exp\Big{(}\frac{-(\log [\psi/\psi_*+1])^2}{2\sigma^2}\Big{)}.
\label{eq:saun}
\end{equation} 
For a fixed set of parameters $(\Phi_*,\psi_*,\alpha)$, the Saunders function is identical to the Schechter function for $\psi \lsim \psi_*$. However, at $\psi>\psi_*$ it cuts off as a Gaussian in log-space with a standard deviation $\sigma$, instead of the sharper exponential cut-off of the Schechter function in real-space. \citet{Salim12} show that Schechter functions typically predict (slightly) fewer galaxies at the largest $\psi$ compared to the actual observations (see e.g. Fig~5 of Smit et al. 2012, and Fig~2 of Salim \& Lee 2012), because of their exponential cut-off at $\psi> \psi_*$. For $\sigma=0.5$ we can boost $dn/d\psi$ by a factor of $\sim$ a few at the high-$\psi$ end, which may help resolve this issue. We repeated our analysis in which we replaced Schechter functions with Saunders functions (using $\sigma=0.5$, and keeping the other parameters fixed), and found that this did not change our results at all. However, this may become more relevant in the future with larger LAE surveys which can probe down to larger Ly$\alpha$ luminosities (and likely larger values of $\psi$).

\subsection{Constraints on Models}
\label{sec:modelcon}

In \S~\ref{sec:intro} we mentioned that empirical constraints on $f_{\rm esc}^{\rm eff}$ may help us understand the basics of Ly$\alpha$ transport in and around galaxies. Our work has several implications:

\begin{itemize}

\item Our best-fit $f_{\rm esc}^{\rm eff}\sim 30-50\%$ at $z=6$. Models that have studied the impact of the IGM on the visibilty of Ly$\alpha$ photons emerging from galaxies at this redshift, consistently conclude that the alone IGM should transmit only $\mathcal{T}_{\rm IGM}\sim 5-30 \%$ \citep[e.g.][]{IGM,Iliev08,Zheng10,Dayal11,Laursen11} of photons through an ionized Universe at $z\sim 6$. Under the reasonable assumption that dust suppresses the emerging Ly$\alpha$ flux by an additional factor, these models would predict effective escape fractions that appear inconsistent with our inferred fraction (and also that of Hayes et al. 2011). A plausible reason for this discrepancy is that the models overestimate the IGM opacity, because they do not include the impact of outflows of optically thick (to Ly$\alpha$ photons) HI gas on the Ly$\alpha$ spectral line profile emerging from galaxies. Winds are known to redshift Ly$\alpha$ photons out of the line resonance, which can strongly increase the fraction of photons transmitted through the IGM \citep[see e.g.][]{DW10}. It is interesting that current constraints on $f_{\rm esc}^{\rm eff}$ provide evidence for winds impacting the Ly$\alpha$ radiative transfer at $z\sim 6$.

\item Our work has also shown that it is possible to reproduce Ly$\alpha$ luminosity functions with a constant ($\psi$-independent) $f_{\rm esc}^{\rm eff}$, in agreement with previous studies \citep[e.g.][]{Nagamine10,Shimizu11}, although we have shown that this applies over a wider range of observed Ly$\alpha$ luminosities (by adding the data from Cassata et al. 2010 to the data from Ouchi et al. 2008 which was used in most previous analyses). We have shown that we `flatten' the predicted luminosity functions by adding a dispersion in $f_{\rm esc}^{\rm eff}$ and/or a `duty cycle' \citep[as in][]{Nagamine10}. This flattening can improve the fit to the observed luminosity function at the bright end. If we flatten the predicted luminosity functions even more (by increasing the dispersion, or reducing the duty cycle), then we need to invoke that $f_{\rm esc}^{\rm eff}$ decreases towards higher $\psi$, which appears to be favored by the observed increase `Ly$\alpha$ fraction' towards fainter drop-out galaxies (see \S~\ref{sec:withscat}).

\end{itemize}

The two points combined appear to favor scenarios in which Ly$\alpha$ photons escape from LAEs through an outflowing ISM. This would explain the large value of $f_{\rm esc}^{\rm eff}$ that has been inferred from the data at $z\sim 6$. Furthermore, a large scatter in $f_{\rm esc}^{\rm eff}$ has been shown to arise naturally in models for LAEs in which Ly$\alpha$ photons scatter through spherically symmetric outflows \citep[][also see Orsi et al. 2012]{Garel12}. Alternatively, the large scatter in $f_{\rm esc}^{\rm eff}$ may also reflect anisotropic escape of Ly$\alpha$ photons from galaxies \citep[as in e.g.][]{Laursen07,Verhamme12}.

\section{Conclusions}
\label{sec:conc}

In this paper, we have constrained the `effective escape' fraction of Ly$\alpha$ photons, $f_{\rm esc}^{\rm eff}$, which is defined as the ratio of the observed Ly$\alpha$ luminosity of a galaxy to its intrinsic Ly$\alpha$ luminosity. This ratio is often referred to in the literature simply as an escape fraction. In \S~\ref{sec:fobsvsfesc} we have argued why we caution against universal usage of the term escape fraction, and why it helps to distinguish between an escape fraction and an effective escape fraction.

We have constrained the effective escape fraction by converting observed star formation functions to observed Ly$\alpha$ luminosity functions. This conversion depends directly on $f_{\rm esc}^{\rm eff}$, and we use observed Ly$\alpha$ luminosity functions at $z=0.35$, $z=3.1$, $z=3.7$, and $z=5.7$ to get constraints on $f_{\rm esc}^{\rm eff}$ at these redshifts. We have explored models in which $f_{\rm esc}^{\rm eff}$ takes on a single value (\S~\ref{sec:noscat}), and in which $f_{\rm esc}^{\rm eff}$ has a dispersion (\S~\ref{sec:withscat}). Models which include a dispersion predict flatter luminosity functions, which appear to be in better agreement with the observations. We note that the flattening predicted by these models cannot be captured by Schechter functions (a Saunders function as in Eq~\ref{eq:saun} would likely be more appropriate).

We found that the effective escape fraction (or its expectation value in a distribution) $f_{\rm esc}^{\rm eff} \sim 1-3\%$ at $z=0$, and that it increases to $f_{\rm esc}^{\rm eff}\sim 10\%$ at $z=3-4$, and to $z=35-50 \%$ at $z=6$ (see Fig~\ref{fig:fig2}). Eq~\ref{eq:fit} provides a convenient fitting formula that encapsulates our main findings. Our results are consistent with previous work \citep[e.g.][]{Hayes10,Blanc,Hayes11}, except at $z\sim 0.35$ where our inferred $f_{\rm esc}^{\rm eff}$ is higher than previous works. We have argued in \S~\ref{sec:results} that this difference may be a result of the systematic uncertainty on $f_{\rm esc}^{\rm eff}$ becoming increasingly large for very small $f_{\rm esc}^{\rm eff}$ in previous analyses. We argued in \S~\ref{sec:modelcon} that our constraint on $f_{\rm esc}^{\rm eff} $ at $z\sim6$ appears higher than predicted by models that do not include winds. This hints at the importance of winds in the Ly$\alpha$ transfer process even at this high redshift.

We have shown that we can reproduce observed Ly$\alpha$ luminosity functions in individual redshift bins with a constant - i.e. independent of $\psi$- $f_{\rm esc}^{\rm eff}$ over up to two orders in $\psi$ and Ly$\alpha$ luminosity (see Fig~\ref{fig:fig2}), in agreement with previous work. We require $f_{\rm esc}^{\rm eff}$ to decrease with $\psi$ - as appears to be favored by observations of drop-out galaxies (see \S~\ref{sec:withscat})- only in models which include a large scatter ($\sigma \gsim 1.0$ dex) in $f_{\rm esc}^{\rm eff}$, or in which star forming galaxies only have a non-zero $f_{\rm esc}^{\rm eff}$ for a fraction $\epsilon_{\rm DC}$ of their life-time and/or a fraction of sightlines (see \S~\ref{sec:modelunc}).

We anticipate that observations of Ly$\alpha$ emitting galaxies in the near future (e.g. with MUSE\footnote{http://www.eso.org/sci/facilities/develop/instruments/muse.html}, Hyper Suprime-Cam\footnote{http://www.naoj.org/Projects/HSC/} and by HETDEX) will determine the Ly$\alpha$ luminosity functions over a wider range of luminosities, and reduce their systematic uncertainties. This may allow for better constraints on $f_{\rm esc}^{\rm eff}$, and its PDF. As illustrated by the discussion in \S~\ref{sec:modelcon}, constraints on the $f_{\rm esc}^{\rm eff}$-PDF yield valuable basic insights into Ly$\alpha$ transfer process on small scales. Perhaps this is more speculative, but the possible dependence of these luminosity functions on the surface brightness threshold of the survey would shed light on the presence of spatially extended Ly$\alpha$ halos around star forming galaxies, which encode valuable information on cold gas around galaxies \citep[e.g][]{Zheng11,DK12,Akila}.

{\bf Acknowledgements} We thank an anonymous referee for helpful, prompt reports which improved the content of this paper significantly, and for insisting we should not use the name we initially had for the quantity now called `effective escape fraction'.

\label{lastpage}

\end{document}